\documentclass[numberedappendix]{emulateapj} 
\newcommand{\bl}[1]{\mbox{\boldmath$ #1 $}}

\slugcomment{Accepted for publication by The Astrophysical Journal}
\shorttitle{The burst mode of accretion}
\shortauthors{Vorobyov \& Basu} 

\begin{document}

\title{The burst mode of accretion and disk fragmentation in the early embedded stages of star formation}
\author{Eduard I. Vorobyov\altaffilmark{1,}\altaffilmark{2} and Shantanu Basu\altaffilmark{3}}
\altaffiltext{1}{Institute for Computational Astrophysics, Saint Mary's University,
Halifax, NS B3H 3C3, Canada; vorobyov@ap.smu.ca.} 
\altaffiltext{2}{Research Institute of Physics, Southern Federal University, Stachki 194, 
Rostov-on-Don, 344090, Russia.} 
\altaffiltext{3}{Department of Physics and Astronomy, The University of Western Ontario, London, ON N6A 3K7, Canada; basu@uwo.ca.}


\begin{abstract}
We revisit our original papers on the burst mode of accretion by incorporating a detailed energy 
balance equation into a thin-disk model for the formation and evolution of circumstellar disks
around low-mass protostars.
Our model includes the effect of radiative cooling, viscous and shock heating, and heating 
due to stellar and background irradiation. Following the collapse from the prestellar phase 
allows us to model the early embedded phase of disk formation and evolution.
During this time, the disk is susceptible to fragmentation, depending upon the properties 
of the initial prestellar core. Globally, we find that higher initial core angular momentum 
and mass content favors more fragmentation, but higher levels of background radiation 
can moderate the tendency to fragment. A higher rate of mass infall onto the disk than that onto the
star is a necessary but not sufficient condition for disk fragmentation.
More locally, both the Toomre $Q$-parameter needs to be below 
a critical value {\it and} the local cooling time needs to be shorter than 
a few times the local dynamical time. Fragments that form during the 
early embedded phase tend to be driven into the inner disk regions,
and likely trigger mass accretion and luminosity bursts that are 
similar in magnitude to FU-Orionis-type or EX-Lupi-like events.
Disk accretion is shown to be an intrinsically variable process, 
thanks to disk fragmentation, nonaxisymmetric structure, and the effect of gravitational torques.
The additional effect of a generic $\alpha$-type viscosity 
acts to reduce burst frequency and accretion variability, and is likely
to not be viable for values of $\alpha$ significantly greater than 0.01.

\end{abstract}


\section{Introduction}

Typical rotation rates of $\sim$ 1 km s$^{-1}$ pc$^{-1}$ $\approx 10^{-14}$ 
rad s$^{-1}$ measured in molecular cloud cores \citep{Goodman93,Caselli} are sufficient to 
provide a significant angular momentum barrier to star formation. Most of the infalling
matter will land on a protostellar disk rather than directly onto a protostar, since magnetic 
braking is rendered ineffective by ohmic dissipation in the near-stellar
environment. Therefore, the early phase of disk formation and evolution holds 
the key to understanding stellar mass accumulation, and sets the initial conditions for a later 
stage of disk evolution during which planets may form by core accretion \citep{Lissauer93}. 
The early disk phase is characterized by episodic accretion, as predicted theoretically in our 
earlier papers \citep{VB05,VB06} and inferred observationally by compiling luminosity
distributions of young stellar objects \citep{Enoch09}. Furthermore, the FU Orionis stars, named 
after the prototype FU Ori \citep{Herbig77}, provide direct evidence of transient
luminosity variations ($3-6$ mag during $\leq 100$ yr). These luminosity bursts have been 
associated with a sharp increase of the mass accretion rate onto the protostar 
\citep{Hartmann96} and various physical mechanisms have been proposed to explain this phenomenon
\citep[see e.g.,][]{Lin85,Bonnell92,Bell94,Armitage01,Lodato04,VB05,VB06,Pfalzner08,Zhu09,Zhu10,
Forgan10,Bate10}.

Our earlier calculations \citep{VB05,VB06,VB07,VB08} have revealed the importance of studying 
disk evolution using a self-consistent method of following the collapse of an 
initial prestellar core. Disk formation occurs after a central stellar core has formed, 
but the disk continues to gain mass from the surrounding infalling envelope. Under certain conditions,
this leads to disk fragmentation and the development of  the ``burst mode''
of accretion, during which fragments are driven onto the protostar and 
episodic high mass accretion events actually account for the majority of mass accumulation onto the
protostar. For example, Fig. 1 of \citet{VB07} and \citet{Vor09} reveals the correlation
of the burst mode with infall from the envelope onto the disk. The above models were characterized 
by a large dynamic range of spatial {\it and} temporal scales, so that core collapse from 
$\sim 10^4$ AU scales down to an inner sink cell of size $5-10$ AU
was resolved, and the evolution followed for up to several Myr
{\it after} the formation of a central protostar.
Aside from the initial discovery of the burst mode, long-term 
evolution revealed that, even after the burst mode ceases, 
the disk settles into a self-regulated mode in which the
Toomre-$Q$ parameter stays near the critical value. In this phase, 
residual accretion due to gravitational torques (resulting from 
persistent low-amplitude nonaxisymmetric structure driven by the 
swing amplifier effect) occurs at a rate that can explain observed
T Tauri star mass accretion \citep{VB07,VB08},  but with late time disk masses
that are about an order of magnitude greater than observational 
estimates \citep{Vor09a} which may really represent lower limits to the actual values
\citep[see e.g.,][]{Andrews07}.
The effect of additional angular momentum transport due to an 
$\alpha$-viscosity was explored by \citet{VB09a,VB09b}, with a finding 
that such an $\alpha$ would likely lie in the range $10^{-3}-10^{-2}$ 
to satisfy observational and theoretical constraints.
An $\alpha$-viscosity in this range could begin to 
dominate gravitational torques only during the late evolution
(Class II, or T Tauri phase) and yield a late time accretion rate
that was a factor $2-3$ greater than that due to gravitational 
torques alone. Values of $\alpha$ well above this range were found to lead
to a lack of accretion variability in the early stages, and quickly lead to very
low mass disks, but with disk lifetimes $\leq 1$ Myr that may be too short.

The above calculations employed a barotropic equation of state, 
which captured the basic
features of the transition from isothermal to polytropic evolution
at high densities (at number density $n \gtrsim 10^{11}$ cm$^{-3}$), as
seen in spherical radiative transfer calculations \citep{Masunaga00}.
However, such calculations 
could not capture the detailed thermodynamics in the vicinity of
forming clumps. That can be a crucial effect in the development of a
clump, including determining whether it can even form at all
\citep{Gammie01,Rice03,Mejia05,Cai08}. Hence, a criticism of the above modeling has been that 
the clump formation and consequent burst mode may not be a robust
result in the case of more realistic thermodynamics. We note that
\citet{VB06} were aware of this difficulty, and tested out models
with high values of polytropic index such that the temperatures 
for densities $n \gtrsim 10^{11}$ exceeded that found in 
radiative transfer calculations. The clump formation and bursts
still occurred in those cases, thanks to significant forcing by 
mass accretion onto the disk during the early phases, although their
number and frequency could be strongly reduced. Bursts were found to be
robust in the context of those models, but their frequency depended
strongly on thermal evolution as well as the mass and angular momentum 
content of infalling material. 

In this paper, we have made a major improvement by implementing the energy
balance equation, which includes radiative cooling, viscous and shock heating,
and heating due to stellar and background irradiation.
 We continue to include angular momentum transport
due to a generic $\alpha$-viscosity term, since mechanisms other than
gravitational torques may be at work.
Numerical resolution is also extended to greater values than 
in our original papers on the burst mode \citep{VB05,VB06}.
An important question is: will the existence of the burst mode be robust 
under these circumstances, and what will be its properties?
We address these questions in Sections 3 - 5. A description of our
model is in Section 2, and we provide extended discussion of the model 
features in Section 6. A summary of results is in Section 7.

\section{Description of the numerical model} 
The main concepts of our numerical approach are explained in detail in \citet{VB06}.
Here we review some main properties and focus mainly on the implementation of  
radiative cooling, viscous and shock heating, 
and heating due to stellar and background irradiation.

We start our numerical integration in the pre-stellar phase, which is 
characterized by a collapsing {\it starless} cloud core, 
continue into the embedded phase of star formation (hereafter, EPSF), during which
a star, disk, and envelope are formed, and terminate our simulations in the T Tauri phase,
when most of the envelope has accreted onto the forming star/disk system.
In the EPSF, the disk occupies the innermost region of our numerical grid, while the 
larger outer part of the grid is taken up by the infalling envelope, 
which is a remnant of the parent cloud core. 
This ensures that the protostellar disk is not isolated in the EPSF but is subject to intense
mass loading from the envelope. In addition, the mass accretion rate onto 
the disk $\dot{M}_{\rm env}$ is not a free parameter of the model 
but is self-consistently determined by the gas dynamics in the envelope.

We introduce a ``sink cell'' at $r_{\rm sc}=6$~AU and impose a free inflow inner boundary condition.
We monitor the gas surface density in the sink cell and 
when its value exceeds a critical value for the transition from 
isothermal to adiabatic evolution, we introduce a central point-mass star.
In the subsequent evolution, 90\% of the gas that crosses the inner boundary 
is assumed to land onto the central star plus the inner axisymmetric disk at $r<6$~AU. 
This inner disk is dynamically inactive, it contributes only to the total gravitational 
potential and is used to secure a smooth 
behavior of the gravity force down to the stellar surface.
The other 10\% of the accreted gas is assumed to be carried away with protostellar jets. 
The latter are triggered only after the formation of the central star.
The fact that we use a sink cell means that our model cannot resolve the formation
of binary (or multiple) stellar (or planetary) systems on spatial scales smaller than the size
of the sink cell.

\subsection{Basic equations}
We make use of the thin-disk approximation to compute the gravitational collapse of rotating, 
gravitationally unstable cloud cores. This approximation is an excellent means to calculate
the evolution for many orbital periods and many model parameters and its justification is provided
in Appendix~\ref{thindisk}. We note that in the thin-disk approximation all material from the 
envelope lands onto the outer disk regions. This is however a reasonable assumption according to \citet{Visser09},
who accurately calculated the gas trajectories in the infalling envelope and found
that the bulk of the infalling material landed onto the disk's outer edge.


The basic equations of mass, momentum, and energy transport in the thin-disk approximation are
\begin{equation}
\label{cont}
\hskip -5 cm \frac{{\partial \Sigma }}{{\partial t}} =  - \nabla_p  \cdot 
\left( \Sigma \bl{v}_p \right),  
\end{equation}
\begin{equation}
\label{mom}
\frac{\partial}{\partial t} \left( \Sigma \bl{v}_p \right) + \left[ \nabla \cdot \left( \Sigma \bl{v_p}
\otimes \bl{v}_p \right) \right]_p =   - \nabla_p {\cal P}  + \Sigma \, \bl{g}_p
+  (\nabla \cdot \mathbf{\Pi})_p, 
\end{equation}
\begin{equation}
\label{energ}
\frac{\partial e}{\partial t} +\nabla_p \cdot \left( e \bl{v}_p \right) = -{\cal P} 
(\nabla_p \cdot \bl{v}_{p}) -\Lambda +\Gamma + \left(\nabla \bl{v}\right)_{pp^\prime}:\Pi_{pp^\prime}, 
\end{equation}
where subscripts $p$ and $p^\prime$ refers to the planar components $(r,\phi)$ 
in polar coordinates, $\Sigma$ is the mass surface density, $e$ is the internal energy per 
surface area, 
${\cal P}=\int^{Z}_{-Z} P dz$ is the vertically integrated
form of the gas pressure $P$, $Z$ is the radially and azimuthally varying vertical scale height
determined in each computational cell using an assumption of local hydrostatic equilibrium \citep{VB09a},
$\bl{v}_{p}=v_r \hat{\bl r}+ v_\phi \hat{\bl \phi}$ is the velocity in the
disk plane, $\bl{g}_{p}=g_r \hat{\bl r} +g_\phi \hat{\bl \phi}$ is the gravitational acceleration 
in the disk plane, and $\nabla_p=\hat{\bl r} \partial / \partial r + \hat{\bl \phi} r^{-1} 
\partial / \partial \phi $ is the gradient along the planar coordinates of the disk. 
The planar components of the divergence of the stress tensor 
$(\nabla \cdot \mathbf{\Pi})_p$, symmetrized velocity gradient tensor $(\nabla \bl{v})_p$, viscous heating
$\left(\nabla \bl{v}\right)_{pp^\prime}:\Pi_{pp^\prime}$, and 
symmetric dyadic $\Sigma \bl{v}_p \otimes \bl{v}_p$ are found
according to the usual rules (see Appendix C).

The gravitational acceleration $\bl{g}_p$ includes the gravity of a central point-mass star 
(when formed), the gravity of an inner disk ($r<r_{\rm sc}$),
and the self-gravity of a circumstellar disk and envelope. The latter component is found 
by solving for the Poisson integral
\begin{eqnarray} 
  \Phi(r,\phi) & = & - G \int_{r_{\rm sc}}^{r_{\rm out}} r^\prime dr^\prime 
                     \nonumber \\ 
      & &       \times \int_0^{2\pi} 
               \frac{\Sigma(r^\prime,\phi^\prime) d\phi^\prime} 
                    {\sqrt{{r^\prime}^2 + r^2 - 2 r r^\prime 
                       \cos(\phi^\prime - \phi) }}  \, ,
\end{eqnarray} 
where $r_{\rm out}$ is the radial position of the computational outer boundary, or,
equivalently, is the initial radius of a cloud core.
This integral is calculated using a FFT technique which applies the 2D Fourier 
convolution theorem for polar coordinates \citep[see][Sect.\ 2.8]{BT87}. 


\subsection{Viscosity}
Viscosity in circumstellar disks may be an important source of mass and angular momentum transport
and heat production. The best candidate to date is turbulent
viscosity induced by the magneto-rotational instability \citep{Balbus91}, though other mechanisms
such as nonlinear hydrodynamic turbulence cannot be completely
eliminated due to the large Reynolds numbers involved. We make no specific assumptions
about the source of turbulence and parameterize
the magnitude of kinematic viscosity using a modified form 
of the $\alpha$-prescription 
\begin{equation}
\nu=\alpha \, c_{\rm s} \, Z \, {\cal F}_{\alpha}(r), 
\end{equation}
where $c_{\rm s}^2=\gamma {\cal P}/\Sigma$ is the square of effective sound speed
calculated at each time step from the model's known ${\cal P}$ and $\Sigma$. The 
function ${\cal F}_{\alpha}(r)=
2 \pi^{-1} \tan^{-1}\left[(r_{\rm d}/r)^{10}\right]$ is a modification to the usual 
$\alpha$-prescription of \citet{SS73} that guarantees that the turbulent viscosity operates 
only in the disk and quickly reduces to zero beyond the disk radius $r_{\rm d}$.
The latter is determined
using a typical density for the disk to envelope transition,
$\Sigma_{\rm d2e}=0.1$~g~cm$^{-2}$, and the radial gas velocity 
\citep[see][for details]{Vor10}.

In this paper, we use a spatially and temporally 
uniform $\alpha$, with its value set to 0.005 in most models. This choice is based on our recent work
\citep{VB09a}, wherein we have studied numerically the secular evolution
of viscous and self-gravitating disks. We found that {\it if} circumstellar disks
around solar-mass protostars could generate and sustain turbulence, then the temporally and 
spatially averaged $\alpha$ should lie in the range $10^{-3}-10^{-2}$. 
Smaller values of $\alpha$ ($\la 10^{-4}$) have little effect on the resultant disk structure and
mass accretion history, which, in this case, is totally controlled by disk gravity.
Larger values ($\alpha \ga 10^{-1}$) destroy circumstellar disks during less than 1.0~Myr of 
evolution and are thus inconsistent with mean disk lifetimes of the order of 2--3~Myr.
Nevertheless, $\alpha$ may vary in time and have greater values in the EPSF (the duration of 
which is usually much shorter than 1~Myr).
The effect of varying $\alpha$ is briefly discussed in Section~\ref{visceffect}.

Viscosity enters the basic equations via the viscous stress tensor $\mathbf{\Pi}$ expressed as
\begin{equation}
\mathbf{\Pi}=2 \Sigma\, \nu \left( \nabla \bl{v} - {1 \over 3} (\nabla \cdot \bl{v}) \mathbf{e} \right),
\label{stressT}
\end{equation}
where $\mathbf{e}$ is the unit tensor. We note that we take no simplifying
assumptions about the form of $\mathbf{\Pi}$ apart from those imposed by the adopted thin-disk 
approximation. 

\subsection{Energy balance}
\label{energy}

Equation~(\ref{energ}) for the internal energy balance includes the usual
compressional term ${\cal P}\left( \nabla_p \cdot \bl{v}_p \right)$, radiative cooling 
$\Lambda$, heating due to stellar/background irradiation $\Gamma$, 
and viscous heating $(\nabla \bl{v})_{pp^\prime}:\Pi_{pp^\prime}$. 
We assume that the heat generated in the disk interior due to viscosity and shocks 
is transported to the disk surface by radiation, which escapes from the disk surface 
at a rate per unit area $2\sigma T_{\rm eff}^4$. This means that we neglect other possible sources 
of heat transport such as convection.
We then make use of the diffusion approximation and link the effective surface temperature
$T_{\rm eff}$ with the midplane temperature of gas $T_{\rm mp}$ via the following relation
$T_{\rm eff}^4=8 T_{\rm mp}^4/(3\tau)$, where $\tau$ is the optical depth \citep{Hubeni90,Johnson03}.
Finally, we substitute $\tau^{-1}$ with $\tau/(1+\tau^2)$ to allow for a smooth transition between
the optically thick and optically thin regimes \citep{Johnson03}. The resulting cooling function is described as
\begin{equation}
\Lambda={\cal F}_{\rm c}\sigma\, T_{\rm mp}^4 \frac{\tau}{1+\tau^2},
\end{equation}
where $\sigma$ is the Stefan-Boltzmann constant 
and ${\cal F}_{\rm c}=2+20\tan^{-1}(\tau)/(3\pi)$ is a function that 
secures a correct transition between the cooling function in the optically thick regime
$\Lambda_{\rm thick}=16\, \sigma \, T_{\rm mp}^4/(3\tau)$ 
and the optically thin one $\Lambda_{\rm thin}=2\,\sigma \,T_{\rm mp}^4\,\tau$. We use 
frequency-integrated opacities of \citet{Bell94}, which are smoothed at the 
principal opacity transitions to allow for iterative solution methods to converge quickly.

Heating due to stellar and background irradiation is treated assuming that this process operates in
the opposite direction to that of radiative cooling, i.e., radiation from the central 
star and natal molecular cloud hits the surface and diffuses down to the midplane where
it transforms into heat.
This allows us to express the heating function as
\begin{equation}
\Gamma={\cal F}_{\rm c}\sigma\, T_{\rm irr}^4 \frac{\tau}{1+\tau^2},
\end{equation}
where $T_{\rm irr}$ is the irradiation temperature at the disk surface 
determined by the stellar and background black-body irradiation as
\begin{equation}
T_{\rm irr}^4=T_{\rm bg}^4+\frac{F_{\rm irr}(r)}{\sigma},
\label{fluxCS}
\end{equation}
where $T_{\rm bg}$ is the uniform background temperature (in our model set to the 
initial temperature of the natal cloud core)
and $F_{\rm irr}(r)$ is the radiation flux (energy per unit time per unit surface area) 
absorbed by the disk surface at radial distance 
$r$ from the central star. The latter quantity is calculated as 
\begin{equation}
F_{\rm irr}(r)= A_{\rm irr}\frac{L_\ast}{4\pi r^2} \cos{\gamma_{\rm irr}},
\end{equation}
where $L_\ast$ is the stellar luminosity, $\gamma_{\rm irr}$ is the incidence angle of 
radiation arriving at the disk surface at radial distance $r$, and $A_{\rm irr}$ is a time-dependent
factor that accounts for the attenuation of stellar radiation in the EPSF 
(see Appendix~\ref{irradiation} for more details).

\begin{figure}
  \resizebox{\hsize}{!}{\includegraphics{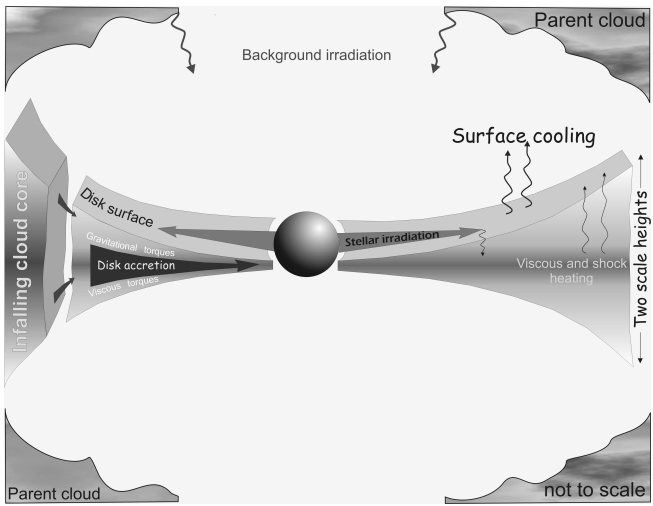}}
      \caption{Schematic representation of the numerical model. See the text for a detailed explanation.}
         \label{fig1}
\end{figure}

The stellar luminosity $L_\ast$ is the sum of the accretion luminosity $L_{\rm accr}=G M_\ast \dot{M}/(2
r_\ast)$ arising from the gravitational energy of accreted gas and
the photospheric luminosity $L_{\rm ph}$ due to gravitational compression and deuterium burning
in the star interior. The stellar mass $M_\ast$ and accretion rate onto the star $\dot{M}$
are determined self-consistently during numerical simulations via the amount of gas passing through
the sink cell. The stellar radius $r_\ast$ is calculated using an approximation formula of \citet{Palla91},
modified to take into account the formation of the first molecular core \citep{Masunaga00}. More specifically,
we assume that during $2\times 10^{4}$~yr after the formation of the 
central protostar, the
stellar radius is $r_\ast=5$~AU. Then, the second atomic core forms and the stellar radius is determined
as
\begin{equation}
r_\ast = \left\{ 
\begin{array}{ll}
2.5~\mathrm{R_\sun} \,\,\, & M_\ast \le 0.4 M_\sun, \\
2.5+4.2(M_\ast-0.4)~\mathrm{R_\sun} \,\,\, & 0.4 M_\sun < M_\ast \le 1.0 M_\sun, \\
5.0~\mathrm{R_\sun} \,\,\, & M_\ast > 1.0 M_\sun.
\end{array} \right.
\label{Andrescheme}
\end{equation}
Transition between these two modes is smoothed over a period of $0.5\times 10^4$~yr.

The photospheric luminosity $L_{\rm ph}$ is taken from the pre-main sequence tracks 
for the low-mass stars and brown dwarfs calculated by \citet{DAntona97}. Unfortunately,
the stellar age in these tracks is difficult to relate with the actual physical
evolution time in numerical simulations of gravitational collapse. A common practice starting from 
\citet{Myers98} is to add $t_{\rm offset}$ to the times of the pre-main sequence tracks 
to account for the delay between the onset of cloud core collapse and the zero-time 
of these tracks. Indeed, after collapse begins, the forming star must wait for some time
before the luminosity due to contraction and deuterium burning (as described by \citet{DAntona97}) 
will begin. The exact value of $t_{\rm offset}$ is however uncertain because it would 
certainly depend on the initial conditions
in a cloud core such as the gas temperature, density enhancement, strength of magnetic fields, etc.

Fortunately, we accurately follow the pre-stellar collapse phase and can actually determine the 
time  $t_{\rm fc}$ that it takes for a cloud core to reach an optically thick density 
of order $10^{11}$~cm$^{-3}$ in its interior and start forming the first (molecular) hydrostatic 
core. 
We then assume that the zero-time of D'Antona \& Mazitelli's tracks 
corresponds to the onset of the formation of the second atomic core $t_{\rm sc}$, 
which follows the formation of the first core after approximately
$2\times 10^{4}$~yr \citep{Masunaga00}. With these assumptions in mind, 
the photospheric luminosity 
$L_{\ast,\rm ph}$ is set to zero for $t<t_{\rm s.c.}=t_{\rm f.c.}+2\times 10^4$~yr and then 
is calculated according to \citet{DAntona97} with the zero-time of their tracks corresponding to
$t_{\rm s.c.}$ in our numerical simulations.
We note that the D'Antona \& Mazitelli's tracks do not cover the very early phases of
stellar evolution. Therefore, we have used a power-law expression
to extrapolate to times earlier than those included in the pre-main sequence tracks, 
$L_{\rm ph} = L_{\rm ph,0}(t/t_0)^4$, where $t_0$ is the earliest time in 
the tracks and $L{\rm ph,0}$ is the pre-main sequence luminosity at this time.


Viscous heating operates in the disk interior and is calculated using the standard expression 
$(\nabla \bl{v})_{pp^\prime} : \mathbf{\Pi}_{pp^\prime}$ (see Appendix~C). We note that 
we use the most general expression for viscous heating and take none
of the popular simplifying assumptions 
(such as disk axisymmetry) apart from those imposed by the thin-disk 
approximation. Heating due to shock waves is taken into account via compressional heating 
${\cal P}\left(\nabla_p \cdot \bl{v}_p \right)$ and 
artificial viscosity. The latter is implemented in the code using the standard prescription 
of \citet{RM57}. 
The gas pressure ${\cal P}$ and internal energy per surface area $e$ are related via 
the ideal gas law 
${\cal P}=(\gamma-1)\, e$, with the ratio of specific heats $\gamma=7/5$.
A more detailed approach would be to implement a variable $\gamma$ as in e.g. \citet{Forgan09}.
However, a rigorous realization of this mechanism requires calculating the excitation levels 
of main atomic and molecular coolants and is beyond the limits of the current paper. We explored
the effect of varying $\gamma$ in our previous paper in the context of polytropic disks \citep{VB06}
and showed that the $\gamma=5/3$ case was usually characterized by disks less prone 
to fragmentation.

In Figure~\ref{fig1} we summarize our model by drawing a schematic picture of the main
model ingredients in the EPSF. A central star is surrounded by a disk which accretes matter
from a collapsing natal cloud core. The infalling material lands onto the disk outer edge and is transported
toward to the inner disk boundary by a combined action of gravitational and viscous torques. 
Mass accretion onto the star, along with stellar compression and deuterium burning, 
give rise to stellar irradiation, part of which is absorbed by the flaring
disk surface  and is transformed into heat in the disk interior. Another source of external heating
is the background irradiation from the natal molecular cloud.
The heat generated in the disk interior by viscosity and shocks is transported to the
disk surface by radiation. The latter escapes from the disk surface giving rise to
the only global cooling mechanism in our model.

\begin{table}
\caption{Model parameters}
\label{table1}
\begin{tabular}{ccccccc}
\hline\hline
Model & $\beta$ & $\Omega_0$ & $r_0$ & $M_{\rm cl}$ & $T_{\rm bg}$ & $\alpha$ \\
\hline
 reference & $1.3\times 10^{-2}$ & 2.7 & 1540  & 0.70 & 10 & $5\times10^{-3}$   \\
 lower-$\beta$  & $2.8\times 10^{-3}$ & 1.2 & 1640  & 0.73 & 10 & $5\times10^{-3}$   \\
 $M_{\rm cl}=0.16$ & $1.3\times 10^{-2}$ & 12 & 340  & 0.16 & 10 & $5\times10^{-3}$ \\
 $M_{\rm cl}=0.23$ & $1.3\times 10^{-2}$ & 8  & 514  & 0.23 & 10 & $5\times10^{-3}$ \\
 $M_{\rm cl}=0.92$ & $1.3\times 10^{-2}$ & 2  & 2060  & 0.92 & 10 & $5\times10^{-3}$ \\
 $T_{\rm bg}=10$ & $1.3\times 10^{-2}$ & 1.8 & 2780  & 1.2 & 10 & $5\times10^{-3}$  \\
 $T_{\rm bg}=20$ & $1.3\times 10^{-2}$ & 4.8 & 1200  & 1.1 & 20  & $5\times10^{-3}$ \\
 $T_{\rm bg}=30$ & $1.2\times 10^{-2}$ & 8.2 & 860  & 1.15 & 30 & $5\times10^{-3}$  \\
 $\alpha=0$ & $1.3\times 10^{-2}$ & 2.7 & 1540  & 0.70 & 10 & 0   \\
 $\alpha=0.05$ & $1.3\times 10^{-2}$ & 2.7 & 1540  & 0.70 & 10 & $5\times10^{-2}$   \\
 \hline
\end{tabular} 
\tablecomments{All masses are in $M_\sun$, temperatures in Kelvin, distances in AU, and angular velocities
in km~s$^{-1}$~pc$^{-1}$.}
\end{table} 
 
\subsection{Solution procedure}
Equations~(\ref{cont})--(\ref{energ}) are solved in polar 
coordinates $(r, \phi)$ on a numerical grid with
$512 \times 512$ grid zones. The radial points are logarithmically spaced.
The innermost grid point is located at the position of the sink cell $r_{\rm sc}=6$~AU, and the 
size of the first adjacent cell varies in the 0.07--0.1~AU range depending on the cloud core 
size.  This corresponds to a radial resolution $\triangle r$=1.1--1.6~AU at 100~AU. 
The outer boundary is reflecting.

We use the method of finite differences with a time-explicit solution procedure similar 
in methodology to the ZEUS code \citep{Stone92}. The advection is treated using the van Leer interpolation scheme.
It is well known that cooling and heating time scales may become much shorter than the dynamical 
time scale, which would result
in prohibitively small time steps. Therefore, the update of the internal energy per surface area 
$e$ due to cooling $\Lambda$ and heating $\Gamma$
is done implicitly using the Newton-Raphson method of root finding, complemented by the bisection method
where the Newton-Raphson iterations  fail to converge. The accuracy is guaranteed by not allowing $e$
to change more than 30\% over one time step. If this condition is violated in a particular cell, 
we employ subcycling for this cell, i.e., the solution is sought with a local time step that
is smaller than the global time step by a factor of 2. The local time step may be further 
decreased until the desired accuracy is reached.

The viscous force and heating terms in Equations~(\ref{mom}) and (\ref{energ}) are 
implemented in the code  using an explicit finite-difference
scheme. This is found to be adequate for $\alpha\la 0.01$ because other terms 
(usually, the azimuthal advection) dominate in 
the Courant condition that controls the time step. However, for higher values of $\alpha$
we find that the viscous terms start to impose strict time step limitations and 
an implicit scheme is desirable in order to extend numerical simulations to 
the Class~II phase of stellar evolution.
A small amount of artificial viscosity is added to the code to smooth out shocks.
The associated artificial viscosity torques integrated over the disk area are negligible
in comparison with gravitational torques. Occasionally, however, the shocks may become strong enough
to impose strict limitations on the Courant condition, which results in a considerable decrease in
the time step of integration. In this case, we use subcycling in the same manner as we do for the
internal energy update due to cooling/heating.


\subsection{Initial conditions}
Initially, cloud cores have surface densities 
$\Sigma$ and angular velocities $\Omega$ typical for a collapsing, axisymmetric, magnetically
supercritical core \citep{Basu97}:
\begin{equation}
\Sigma={r_0 \Sigma_0 \over \sqrt{r^2+r_0^2}}\:,
\label{dens}
\end{equation}
\begin{equation}
\Omega=2\Omega_0 \left( {r_0\over r}\right)^2 \left[\sqrt{1+\left({r\over r_0}\right)^2
} -1\right],
\label{omega}
\end{equation}
where $\Omega_0$ is the central angular velocity and 
$r_0$ is the radius of central near-constant-density plateau defined 
as $r_0 = \sqrt{A} c_{\rm s}^2 /(\pi G\Sigma_0)$. 
We note that the above form of the column density is very similar
to the integrated column density of a Bonnor-Ebert sphere
\citep{Dapp09}.  
Furthermore, equation~(\ref{dens}) at large radii $r\gg r_0$ leads to 
the gas volume density distribution
$\rho =A c_{\rm s}^2/(2 \pi G r^2)$, if it is
integrated in the vertical direction assuming a local vertical hydrostatic equilibrium, 
i.e., $\rho = \Sigma/(2Z)$ and $Z=c_{\rm s}^2/(\pi G \Sigma)$. This means that our initial 
gas surface density configuration can be considered to have a factor of $A$ 
positive density enhancement compared to
that of the singular isothermal sphere $\rho_{\rm SIS} =c_{\rm s}^2/(2\pi G r^2)$ \citep{Shu77}.
Throughout the paper, we use $A=1.2$.

Cloud cores are also characterized by the ratio of
rotational to gravitational energy $\beta=E_{\rm rot}/|E_{\rm grav}|$, where the
rotational and gravitational energies are calculated as
\begin{equation}
E_{\rm rot}= 2 \pi \int \limits_{r_{\rm sc}}^{r_{\rm
out}} r a_{\rm c} \Sigma \, r \, dr, \,\,\,\,\,\
E_{\rm grav}= - 2\pi \int \limits_{r_{\rm sc}}^{\rm r_{\rm out}} r
g_r \Sigma \, r \, dr.
\label{rotgraven}
\end{equation}
Here, $a_{\rm c} = \Omega^2 r$ is the centrifugal acceleration, and $r_{\rm out}$
is the outer cloud core radius. The adopted values of $\beta$ lie within
the limits inferred by \citet{Caselli} for dense molecular cloud cores, $\beta=(10^{-4} - 0.07)$.
Cloud cores are initially isothermal, with the uniform gas temperature 
taking values between $T_{\rm init}=10$~K and 30~K, depending on the model.
In addition, every model core is characterized by a distinct ratio $r_{\rm out}/r_0=6$ 
in order to generate gravitationally unstable truncated cores of similar form.

For the in-depth analysis, we consider a model with 
$M_{\rm cl}=0.7~M_\sun$, $\beta=1.3\times 10^{-2}$, $A=1.2$, $\alpha=5\times 10^{-3}$ 
(the viscous $\alpha$-parameter), and $T_{\rm init}=10$~K. These and other model parameters 
are summarized in Table~\ref{table1}. 
This model (hereafter, the reference model) is chosen solely because it best represents 
the main characteristics of disk fragmentation in the embedded phase of star 
formation. Other models will be introduced as the need arises.


\section{Gravitational instability and disk fragmentation}
\label{fragment}

\begin{figure*}
 \centering
  \includegraphics[width=14cm]{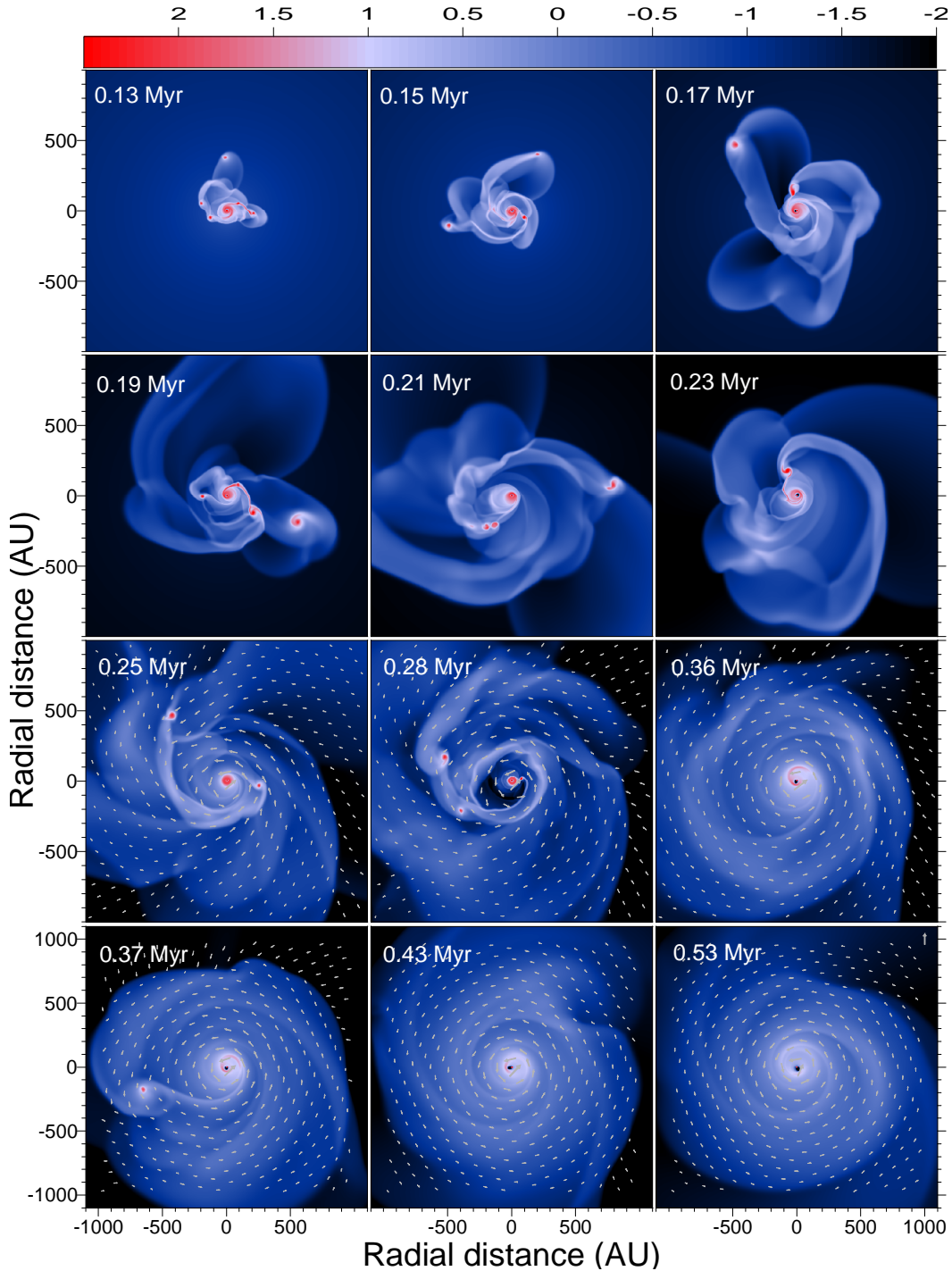}
      \caption{Gas surface density distribution (g cm$^{-2}$, log units) in the reference model at several
       time instances after the formation of the central star (located in the coordinate center). Two
      bottom rows also shows a gas velocity field superimposed onto the 
      surface density distribution. 
      The vertical arrow in the bottom-right panel has a dimension of 5~km~s$^{-1}$.}
         \label{fig2}
\end{figure*}
Theoretical and numerical studies of the evolution of protostellar disks indicate that 
disk fragmentation is a complicated phenomenon, which can be influenced by both the internal disk physics
and external environment. The latter may influence the disk susceptibility to fragmentation 
directly (through, e.g., disk
irradiation) or indirectly by setting the initial conditions in cloud cores that favor or 
disfavor fragmentation in subsequently formed disks.
The following four criteria for disk fragmentation are best studied and their significance is
well established.
\begin{enumerate}
\item The ratio of the local cooling time $t_{\rm c}=e/\Lambda$ to the local dynamical 
time $\Omega^{-1}$ is smaller than a few, i.e., $t_{\rm c} \Omega
\le C$ \citep{Gammie01,Rice03,Mejia05}. The actual value of $C$ may vary 
depending on the physical conditions in the disk, e.g., $C$ may depend on the disk thickness, chemical
composition, dust content, etc. In the following text, we will refer to the dimensionless 
quantity $t_{\rm c} \Omega$ as the ${\cal G}$-parameter and adopt ${\cal G}=1$ as a fiducial
critical value.

\item The Toomre criterion $Q=c_{\rm s} \Omega/(\pi G \Sigma)$ for a Keplerian disk 
is smaller than some critical value $Q_{\rm cr}$, usually taken to be 
unity \citep{Toomre64}. Here again, $Q_{\rm cr}$ may depend on the physical conditions and
may vary by a factor of unity. The Toomre criterion implies that the gas surface 
density $\Sigma$ should be sufficiently high for a disk to fragment. 
 This criterion, along with ${\cal G}\le$~1--3, is often invoked when analyzing the disk susceptibility
to fragmentation \citep[e.g.][]{Rafikov05}.
There is, however, a catch---too high $\Sigma$ may prevent 
fragmentation due to increased opacity and cooling time \citep{Nero09}. 
In other words, there exists minimum and maximum values of $\Sigma$ between which 
the instability and fragmentation are expected to occur.
This means that any numerical simulation that
starts from a {\it pre-defined} star/disk system with some disk-to-star mass ratio
may run the risk of not revealing disk fragmentation if the initial $\Sigma$ is too high. 
This is generally not a problem in numerical simulations that form disks self-consistently (such as
our own), because during the disk formation phase $\Sigma$ naturally increases 
from low toward higher values and the disk may pass through the unstable phase.

\item 
The amount of rotation in the natal cloud core should be sufficiently large in order
to form extended and massive protostellar disks \citep{VB06,Kratter08,Vor09,Rice10,Machida10}.

\item The time-averaged rate of mass accretion onto the disk $\langle\dot{M}_{\rm d}\rangle$ 
is greater than the time-averaged mass accretion onto the star $\langle \dot{M}_\ast \rangle $
so that $\Sigma$ quickly increases with time and may reach the unstable regime 
\citep{VB06,VB07,Vor09,Kratter10,Boley09a}.

\end{enumerate}

We analyze the significance of these four criteria for disk fragmentation using 
our reference model.
Figure~\ref{fig2} shows a series of images of the gas surface density 
(in g~cm$^{-2}$, log units) in the inner
1000~AU at different times since the formation of the central star. The disk begins to form 
at $t\approx 0.08$~Myr and by $t=0.13$~Myr a well-developed spiral pattern and several
dense clumps are clearly visible. The clumps are almost always located in the spiral arms, 
suggesting that they form via fragmentation of the densest and coldest arms.
Most fragments, however, do not live long. They are driven into the disk inner regions and 
through the sink cell (and probably
onto the star) but other fragments take their place. Some of them are massive enough to host mini-disks
of their own. Typical fragment masses lie in a wide range from several Jovian masses 
to low- and intermediate mass brown dwarfs. The mass spectrum of the fragments depends on the disk 
and cloud core properties and may vary from model to model.

The disk slowly grows in mass and size due to mass loading from the envelope 
(most of which is off the spatial scale in Figure~\ref{fig2}).
The disk structure is rather irregular, particularly in the early evolution. 
The gas velocity field in the bottom rows of Figure~\ref{fig2} reveals large
non-circular motions, contractions, and expansions caused by ongoing angular 
momentum redistribution between the fragments and the rest of the disk (in particular, between
the fragments and spiral arms). The disk in this early phase of evolution is most 
certainly {\it not} in a steady state and approximating the early disk evolution 
using a steady-state concept may be misleading.


\begin{figure}
 \centering
  \resizebox{\hsize}{!}{\includegraphics{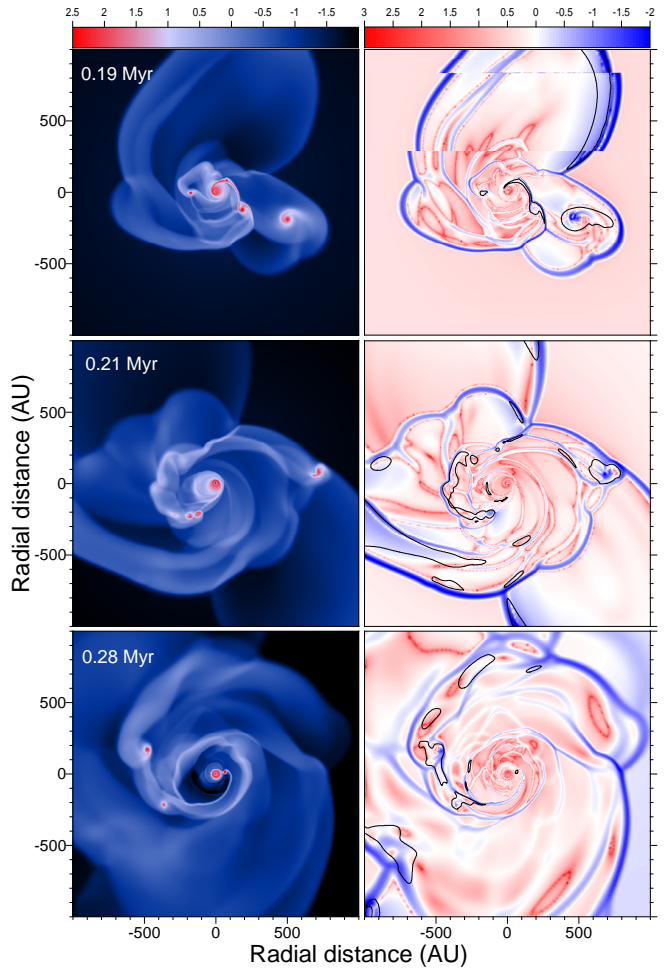}}
      \caption{Gas surface density distribution (left column, g cm$^{-2}$, log units) and
      the spatial distribution of the ${\cal G}$-parameter (right column, log units) 
      in the reference model at three
      typical times after the formation of the central star. Red contour lines comprise 
      gravitationally unstable regions according to the Toomre criterion, $Q<1$. Fragmentation
      is supposed to occur in the regions where ${\cal G}<1$ and $Q<1$ simultaneously.}
         \label{fig3}
\end{figure}

Figure~\ref{fig2} reveals that the disk in the reference model is readily susceptible to fragmentation
in the early evolution. How does the model comply with the four fragmentation criteria outlined above?
Are all four conditions satisfied?
We start with examining the importance of criteria~1 and 2 and search for any disk regions 
that are simultaneously characterized by both $Q<1$ and ${\cal G}<1$.
Figure~\ref{fig3} presents several typical gas surface density distributions
(left column, g cm$^{-2}$, log units) and the spatial distribution of the 
corresponding ${\cal G}$-parameter  
(right column, log units). In the latter case, those regions that cool sufficiently fast 
for fragmentation to take place (${\cal G}<1$) are plotted with blue, while slowly cooling regions
with ${\cal G}>1$ are plotted with red. Disk regions shown with white are near the border of stability,
${\cal G}=1$.
The black contour 
lines delineate the regions of the disk that are prone to fragmentation according to the Toomre criterion,
$Q<1$. It is clearly seen that {\it there
are} regions in the disk where the first two criteria for fragmentation, 
${\cal G}\equiv t_{\rm c}\Omega<1$ and $Q<1$,
are fulfilled simultaneously. These are the fragments, 
especially those located in the outer disk regions, and certain parts of the spiral arms.
It is seen that favorable sites for fragmentation lie preferably at large radii, 
implying that many fragments form at $r\ga100$~AU from the star but 
are driven later in the inner regions via exchange of angular
momentum with the disk and, especially, with the spiral arms. This migration phenomenon\footnote{The
animation of this migration process can be downloaded at www.ap.smu.ca/$\sim$vorobyov/}
was demonstrated by us in the context of barotropic disks \citep{VB06}.

\begin{figure}
 \centering
  \resizebox{\hsize}{!}{\includegraphics{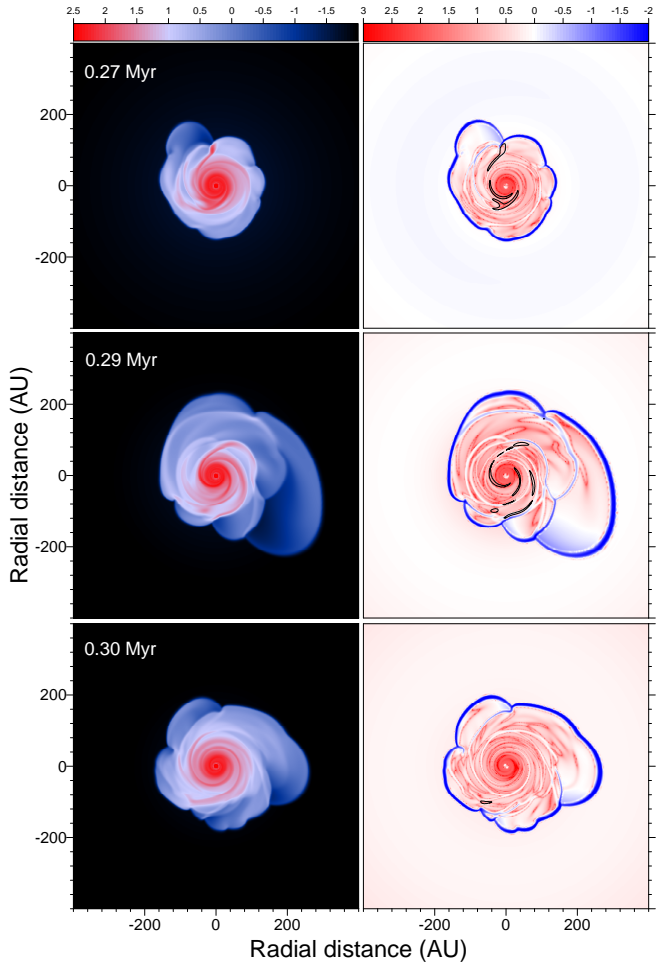}}
      \caption{Same as in Figure~\ref{fig3} but for the lower-$\beta$ model, $\beta=2.8\times 10^{-3}$.
      Note the lack of fragmentation.}
         \label{fig4}
\end{figure}

Figure~\ref{fig3} demonstrates that criteria~1 and 2 for disk fragmentation are fulfilled
in the reference model. What about the other two criteria?
Criterion~3 is essentially an {\it initial condition} imposed on the cloud core 
which states that the rate of cloud core rotation should be sufficiently high 
for disk fragmentation to take place. 
To investigate the importance of this condition, we consider another model that is similar to the
reference model but has a smaller initial rotation rate (hereafter, lower-$\beta$ model). 
In particular,  we set the ratio of rotational
to gravitational energy to $\beta=2.8\times 10^{-3}$ (in contrast to $\beta=1.3\times 10^{-2}$
in the reference model) by decreasing the value of $\Omega_0$ in Equation~(\ref{omega}).
The resulting distributions of the gas surface density  $\Sigma$, ${\cal G}$-parameter, and
Toomre parameter $Q$ are shown in Figure~\ref{fig4}. The layout of the figure is the same as that
of Figure~\ref{fig3} but the spatial scale is different. It is evident that the lower-$\beta$ 
model has no well-defined fragments,
though some transient density enhancements within the spiral arms are visible.
The lack of disk fragmentation is not surprising---there are hardly any regions 
in the disk where criteria~1 and 2 are satisfied simultaneously. In fact, by $t=0.3$~Myr, the disk
lacks regions with $Q<1$ and regions with ${\cal G}<1$ are mostly located near the disk
outer edge where intense cooling of the shocked gas (due to accretion from the envelope) takes place.

There are two major factors that work against disk fragmentation in the lower-$\beta$ model.
First, the disk size is considerably smaller than that of the reference model due to 
a smaller centrifugal radius $r_{\rm cf}=\Omega^2 r^4/GM(r)$. Smaller disks 
are subject to a stronger stabilizing influence of stellar irradiation.
Second, the disk mass in the lower-$\beta$ model is on average 20\% smaller than
that of the reference model, which also increases the disk stability against 
fragmentation in the lower-$\beta$ model by raising the value of $Q$.
In addition, small disks may be optically thick and thus cooling too slow
to fragment \citep[e.g.][]{Rice09,Clarke09}.

The above analysis indicates that the initial conditions in a natal cloud core (in particular, 
the amount of rotation), are of considerable importance for the future disk evolution. 
In models with low $\beta$, the resulting disks are unlikely to fragment due to small disk sizes
and masses. In this sense, 
criterion~3 is a necessary condition for disk fragmentation but not a sufficient one.
As will be demonstrated later, disk propensity to fragment
also depends on other factors such as magnetic fields, initial cloud core temperature and mass, 
etc. In this context, it is difficult to provide reliable estimates as to the exact amount 
of rotational energy (as specified, for example, by the ratio $\beta$ of the 
rotational to gravitational energy) that a cloud core needs in order 
to produce disks capable for fragmentation. 
Therefore, we believe that providing any critical values of $\beta$ for disk fragmentation may be 
misleading unless exact initial conditions in cloud cores are specified.

In the following section, we will consider mass accretion rates onto the disk and the star and 
discuss the significance of criterion~4 for disk fragmentation.

\section{The burst mode of accretion}
\label{burst}
Fragments that form in the disk pass through the sink cell as they migrate into 
the inner disk via exchange of angular momentum with the spiral arms.
The ultimate fate of these fragments is uncertain and is largely dependent on how quickly they can 
contract from their initial size of several AU to a planetary size to avoid tidal destruction.
The contraction time for a Jupiter-mass clump to reach a central temperature of 2000 K,
i.e., the temperature required to dissociate H$_2$ to trigger rapid
collapse, may be as long as $\mathrm{a~few}~\times10^5$ yr \citep{Helled06}. Considering 
a fast timescale of inward radial migration in the embedded phase---a few tens of orbital
periods---we believe that most of these fragments\footnote{The most massive fragments may survive 
and form giant planets or brown dwarfs on close orbits.}
will be tidally destroyed when approaching the central star, 
thus converting its gravitational energy to the accretion luminosity
and producing an FU-Ori-like luminosity burst. This phenomenon is called the burst mode of accretion
and it has been extensively studied by us for the case of barotropic disks \citep{VB05,VB06}. Here,
we confirm that a more accurate treatment of disk thermodynamics does not qualitatively affect
our earlier conclusions. 
However, as our recent simulations of barotropic disks 
have shown, some of the fragments that form in the late embedded phase may survive and
evolve eventually into giant planets on wide orbits \citep{VB10}.

\begin{figure}
  \resizebox{\hsize}{!}{\includegraphics{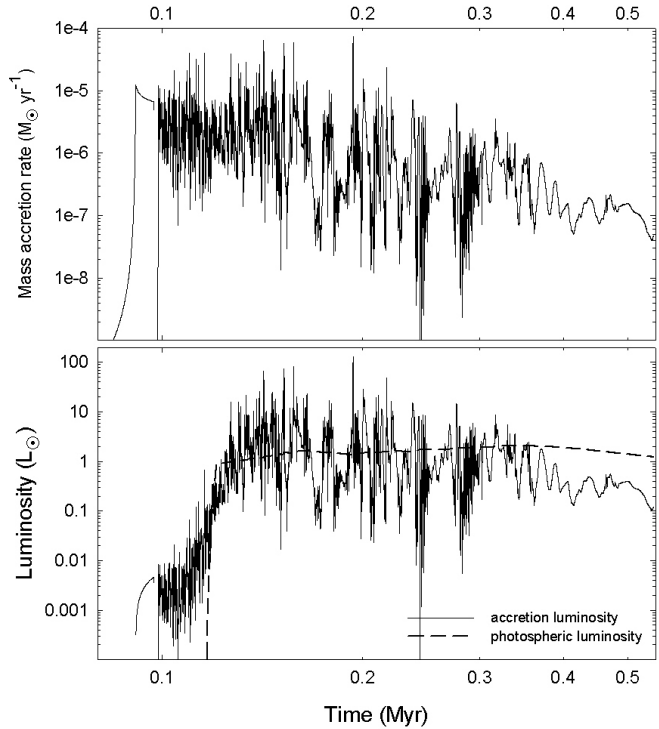}}
      \caption{Mass accretion rate onto the star (top) and stellar luminosity (bottom) as a function
      of time elapsed since the beginning of collapse in
      the reference model. In particular, the solid and dashed lines in the bottom
      panel show the accretion and photospheric luminosities, respectively.}
         \label{fig5}
\end{figure}

The instantaneous mass accretion rate from the disk onto the star
$\dot{M}_\ast$ is found in our model as the mass passing through the sink cell
per one time step of integration (which in physical units is usually equal to 10--20 days). 
We also calculate the instantaneous mass accretion rate onto the disk from 
the infalling envelope $\dot{M}_{\rm d}$ as the mass passing (per one time step of integration)
through a radial annulus located just outside the disk outer edge.
Figure~\ref{fig5} presents the time evolution of the mass accretion rates and luminosities 
in the reference model. In particular, the top panel shows $\dot{M}_\ast$, while 
the bottom panel---accretion luminosity $L_{\rm accr}$ (solid line) 
and photospheric luminosity $L_{\rm ph}$ (dashed line).

In the pre-stellar phase, $\dot{M}_\ast$ is negligible but quickly rises to $\approx 10^{-5}~M_\sun$~yr$^{-1}$
when the gas volume density in the sink cell exceeds $10^{11}$~cm$^{-3}$ and a central stellar core
begins to form at $t\approx0.08$~Myr after the onset of collapse.
The subsequent short period of near-constant accretion corresponds to the phase
when the infalling envelope lands directly onto the forming star\footnote{In fact, this period is 
expected to be even shorter since $r_{\rm sc} \gg r_\ast$.}. A sharp drop in
$\dot{M}_\ast$ follows shortly, which manifests the beginning of the disk formation phase. Subsequently,
the infalling envelope accretes onto the forming disk rather than directly onto the star. 
This transient drop
in $\dot{M}$ occurs due to the fact that the disk mass is initially too small to 
drive a substantial accretion rate onto the star either due to viscous or gravitational torques.
As the evolution proceeds, the disk accretes mass from the infalling envelope and a qualitatively
new phase of mass accretion ensues, in which $\dot{M}_\ast$ shows variability by several orders of 
magnitude. Short episodes of high-rate accretion (caused by the passage of disk fragments 
through the sink cell) are followed by longer periods of low-rate accretion 
(caused by a temporary disk expansion and stabilization). This highly variable accretion makes
the star sporadically increase its {\it total} luminosity, as illustrated in the bottom
panel of Figure~\ref{fig5}. Several clear-cut luminosity outbursts with $L_{\rm accr}$
as high as $100~L_\sun$ and many more weaker bursts (solid line) are evident 
against the background of a near-constant photospheric luminosity with 
$L_{\rm ph} \sim 1.0~L_\sun$ (dashed line). The stronger bursts may represent FU~Orionis-like 
eruptions, typical for the early evolution of a protostar, 
while weaker ones may manifest EX~Lupi-like eruptions (EXors), typical for the later evolution.
We note that the exact time for the onset of the photospheric luminosity is rather uncertain and 
may shift to later times (see discussion in Section~\ref{discuss}), which would result 
in the early luminosity bursts being considerably stronger in amplitude.

\begin{figure}
  \resizebox{\hsize}{!}{\includegraphics{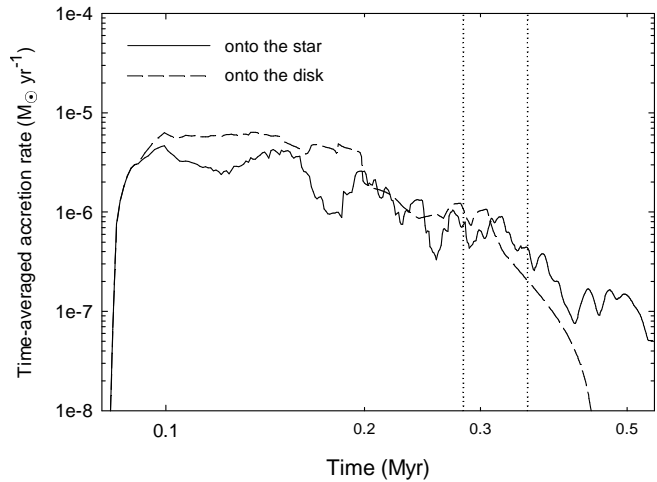}}
      \caption{Time-averaged accretion rates onto the star (solid line) and onto the disk (dashed line)
      versus time since the onset of collapse in the reference model.
      The vertical dotted lines mark the onset of the Class~II
      phase as inferred from the ratio of the envelope mass to the initial cloud core mass, 0.1 and
      0.05 for the left and right lines, respectively}
         \label{fig6}
\end{figure}

We can now verify if our reference model complies with criterion~4 for disk fragmentation
outlined in the previous section. This criterion requires that the rate of mass accretion onto the disk
$\dot{M}_{\rm d}$ be {\it on average} greater than that onto the star $\dot{M}_\ast$. 
Figure~\ref{fig6} presents the time-averaged mass accretion rates onto the star $\langle\dot{M}_\ast\rangle$
(solid line) and onto the disk $\langle \dot{M}_{\rm d}\rangle$ (dashed line) as a function of time
since the beginning of collapse. The averaging is done over a period
of $15000$~yr. In the early evolution ($t<0.2$~Myr), $\langle \dot{M}_{\rm d}\rangle$ is systematically
greater than $\langle\dot{M}_\ast\rangle$ and this phase
is characterized by the strongest burst activity. In the subsequent time period between 0.2~Myr 
and 0.3~Myr, both time-averaged accretion rates are of similar magnitude and the burst phenomenon 
persists, though with somewhat lesser frequency and amplitude. After $t=0.4$~Myr, 
$\langle \dot{M}_{\rm d}\rangle$ becomes systematically lower than $\langle \dot{M}_\ast \rangle$
and  the burst activity in this late phase diminishes. However,
some small variations in $\dot{M}_\ast$ persist even to later times.

Let us define the end of the embedded phase and the onset of the Class~II phase of star formation 
as the time when the envelope empties, and its mass $M_{\rm env}$ drops below 5--10\% of 
the initial cloud core mass $M_{\rm cl}$.
The vertical lines in Figure~\ref{fig6} correspond to the evolution times when $M_{\rm env}/M_{\rm cl}=0.1$
(left) and $M_{\rm env}/M_{\rm cl}=0.05$ (right). It is seen that  $\langle\dot{M}_{\rm d}\rangle \ge
\langle\dot{M}_\ast\rangle$ in the Class~0 and I phases, while $\langle\dot{M}_{\rm d}\rangle <
\langle\dot{M}_\ast\rangle$ in the Class~II phase. Hence, disk fragmentation and the associated 
burst phenomenon are likely to take place in {\it the embedded phase of star formation}, but are
unlikely later in the evolution simply because mass loading from the envelope diminishes in this phase.

\begin{figure}
  \resizebox{\hsize}{!}{\includegraphics{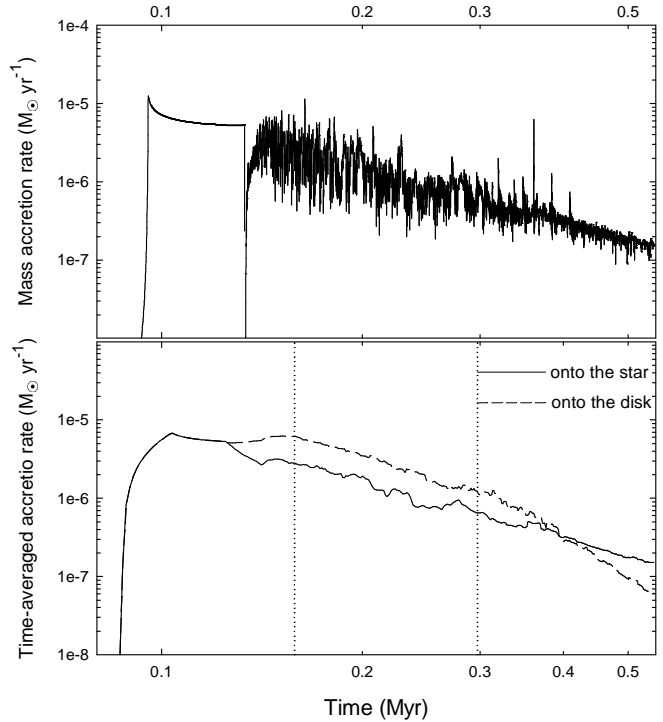}}
      \caption{Instantaneous mass accretion rate (top) and time-averaged mass accretion rates
      (bottom) in the lower-$\beta$ model, which is characterized by a five times lower
      $\beta=2.8\times 10^{-3}$ than that of the reference model. In particular, the 
      solid and dashed lines in the 
      bottom panel show
      the time-averaged accretion rates onto the star and onto the disk, respectively. Vertical dotted
      lines mark the onset of the Class~I (left) and Class~II (right) phases. }
         \label{fig7}
\end{figure}

Figure~\ref{fig6} demonstrates that criterion~4 is fulfilled in the reference model.
Is this criterion sufficient for disk fragmentation to take place? 
In Figure~\ref{fig7} we present the instantaneous accretion rates $\dot{M}_\ast$ (top panel)
and time-averaged accretion rates (bottom panel) in the lower-$\beta$ model 
introduced in the previous section. This model has a (roughly) five times smaller value of 
$\beta=2.8\times 10^{-3}$ as compared to that of the reference model and shows hardly any signs 
of disk fragmentation 
(see Figure~\ref{fig4}). The lack of disk fragmentation manifests itself by a considerably 
weaker accretion variability than in the reference model---there are only order-of-magnitude flickering
in $\dot{M}_\ast$ and one moderate accretion burst. 
However, when we turn to the time-averaged accretion rates (bottom panel), we see that $\langle \dot{M}_{\rm
env}\rangle$ (dashed line) is actually greater than $\langle \dot{M}_\ast \rangle$ (solid line) 
in the EPSF, indicating that criterion~4 for disk fragmentation is fulfilled
in the lower-$\beta$ model.  This example convincingly demonstrates that the fulfillment of
criterion~4 is necessary but not sufficient for disk fragmentation to occur. The disk mass and 
radius in the lower-$\beta$ model seem to be too small even in the case of a strong mass 
loading from the envelope.

\section{The effect of initial conditions on the burst mode of accretion}
In Section~\ref{fragment}, we have already demonstrated the importance of rotation  
for the development of the burst mode of accretion in the early phases of stellar evolution.
In this section, we study the effect that other initial conditions in collapsing cloud cores
(such as cloud core mass and temperature, magnetic fields, etc.)
may have on the strength and frequency of the bursts. 

\subsection{Initial cloud core mass}
\label{coremass}
There is at least one good reason to believe that the initial mass of a cloud core 
should have a significant effect on the subsequent disk evolution---more massive cloud cores
are expected to form more massive disks. This is simply because more massive cloud cores
have larger sizes\footnote{A cloud core may also increase its mass via 
density enhancement.}  and, as a consequence, 
larger centrifugal radii $r_{\rm cf}$ for any reasonable radial mass 
distribution. Hence, we can expect disks formed from more massive cloud cores to have 
a higher tendency for fragmentation and a stronger accretion variability. This effect has been 
confirmed in the context of barotropic disks \citep{VB06,Vor09}. A 
similar tendency was demonstrated by \citet{Kratter08}, who showed that stars of greater
mass tend to have disks that are more susceptible to fragmentation.

\begin{figure}
  \resizebox{\hsize}{!}{\includegraphics{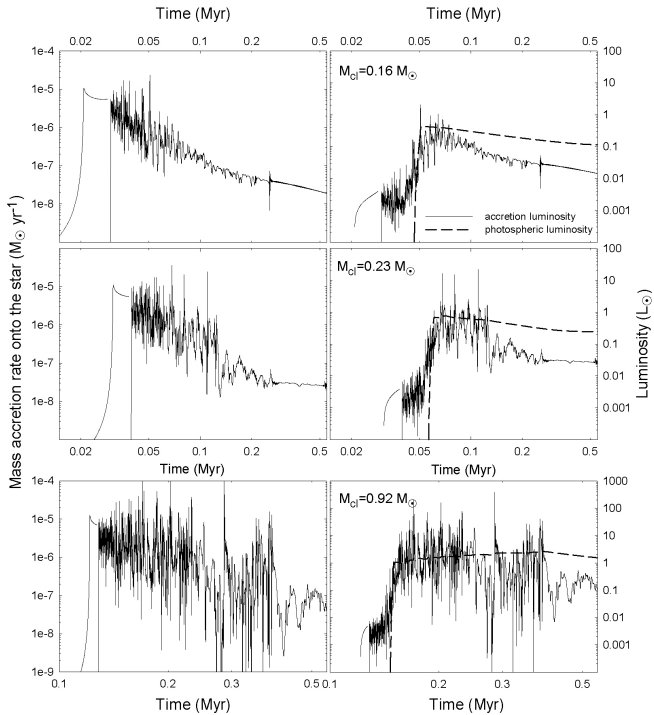}}
      \caption{Mass accretion rate onto the star (left column) and stellar luminosity (right column)
      as a function of time since the beginning of collapse
      in the $M_{\rm cl}=0.16~M_\sun$ model (top), $M_{\rm cl}=0.23~M_\sun$ model (middle),
      and $M_{\rm cl}=0.92~M_\sun$ model (bottom). In particular, the solid and dashed lines 
      in the bottom panel show the accretion and photospheric luminosities, respectively. }
         \label{fig8}
\end{figure}

Figure~\ref{fig8} presents the mass accretion rates onto the star (left column) and
accretion and photospheric luminosities (right column) in three models with $M_{\rm cl}=0.16~M_\sun$
(top row), $M_{\rm cl}=0.23~M_\sun$ (middle row), and $M_{\rm cl}=0.92~M_\sun$ (bottom row). 
In the following text, we refer to these models as the $M_{\rm cl}=0.16~M_\sun$ model,
$M_{\rm cl}=0.23~M_\sun$ model, and $M_{\rm cl}=0.92~M_\sun$ model, respectively.
Other parameters of these models are identical to the parameters of the reference model 
and are summarized in Table~\ref{table1}. It is seen that models with lower $M_{\rm
cl}$ are characterized by a lower accretion variability, suggesting that the 
disk propensity to fragment declines with decreasing cloud core mass. 
The $M_{\rm cl}=0.16~M_\sun$ model exhibits 
hardly any (or very weak) accretion and luminosity bursts, with the photospheric luminosity dominating
the total radiation flux for most of the evolution. The burst mode becomes prominent in 
the $M_{\rm cl}=0.23~M_\sun$ model, which shows three well-defined luminosity outbursts.
As the cloud core mass continues to increase, the burst frequency and intensity also increase and 
the $M_{\rm cl}=0.92~M_\sun$ model demonstrates multiple luminosity outbursts with $L_{\rm accr}
\sim$~10--100~$L_\sun$ and several ones with $L_{\rm accr} > 100~L_\sun$, indicating the onset
of vigorous gravitational instability and disk fragmentation.

We point out that all three models have the same value of $\beta=1.3\times 10^{-2}$, yet the
$M_{\rm cl}=0.16~M_\odot$ and $M_{\rm cl}=0.23~M_\odot$ models have a considerably weaker burst activity
than the $M_{\rm cl}=0.92~M_\odot$ model. This example demonstrates the importance of the 
initial cloud core mass for the development of disk fragmentation and 
associated burst mode of accretion. For disk fragmentation to take place, it is not sufficient 
for a cloud core to have a high initial rate of rotation---the initial cloud core mass should 
also be sufficiently high.
We also note that as $M_{\rm cl}$ increases in Figure~\ref{fig8}, the resulting total luminosity 
also increases but this does not suppress disk fragmentation. 
The growing disk mass outweighs the stabilizing influence of stellar irradiation, 
at least for stars with $M_\ast\la1.0~M_\odot$. Our conclusion is in line with
that of \citet{Rice10} who argue that the primary requirement for disk fragmentation is
large enough $\beta$ to produce disks with radii large enough for fragmentation. Indeed, we may 
form disk of greater size not only by increasing $\beta$ but also by taking 
a larger (and hence more massive) cloud core.

\subsection{Higher initial cloud core temperature}
In the reference model, we set the initial cloud core temperature to $T_{\rm init}=10$~K. 
According to our model assumptions, this value is physically determined by the temperature 
of the background blackbody radiation  $T_{\rm bg}$, i.e., $T_{\rm init}=T_{\rm bg}$.  
However, $T_{\rm bg}$ may be higher and this may influence the subsequent evolution 
of the cloud core in at least three ways. First, the rate of mass accretion onto the disk will
be greater because $\dot{M}_{\rm d}$ is proportional to the cube of the sound speed. 
This effect will assist disk fragmentation. 
Second, the background radiation flux  will grow and moderate the disk tendency to fragment
by systematically increasing the disk temperature \citep{Cai08}.
And lastly, an increased rate of mass accretion onto the disk may eventually lead to an increased
rate of mass accretion onto the star, thus raising the accretion luminosity and
contributing to another factor against disk fragmentation.
It is unclear a priori which of the three key factors would dominate the disk evolution.

\begin{figure}
  \resizebox{\hsize}{!}{\includegraphics{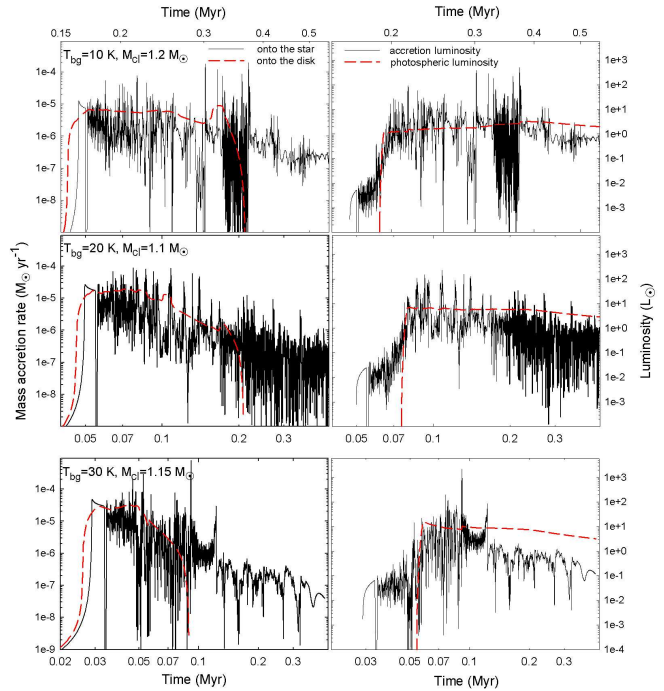}}
      \caption{Mass accretion rates (left column) and stellar luminosity (right column)
      as a function of time since the beginning of collapse
      in the $T_{\rm bg}$=10~K model (top), $T_{\rm bg}$=20~K model (middle),
      and $T_{\rm bg}$=30~K model (bottom). In particular, the black solid and red dashed lines 
      in the left column
      present the instantaneous mass accretion rate onto the star and time-averaged mass accretion rate
      onto the disk, respectively, while these lines 
      in the right column show the accretion and photospheric luminosities, respectively. 
      A color version of this figure is available in the online journal.}
         \label{fig9}
\end{figure}

To study the effect of varying background temperature, 
we consider three models that have similar cloud core masses and rotation rates but different 
background temperatures: $T_{\rm bg}=$10~K, $T_{\rm bg}$=20~K, and $T_{\rm bg}$=30~K. In the following
text, we refer to these models as the $T_{\rm bg}=10$~K model, $T_{\rm bg}=20$~K model, and $T_{\rm bg}=30$~K model, respectively, and their parameters are listed in Table~\ref{table1}.
We specifically choose models with similar $M_{\rm cl}$ and $\beta$ in order to avoid any possible interference
with the effects based on different cloud core masses and rotation rates considered in Section~\ref{burst}
and \ref{coremass}, respectively.
Figure~\ref{fig9} presents mass accretion rates (left column) and
luminosities (right column) as a function of time since the onset of gravitational
collapse in the $T_{\rm bg}$=10~K model (top row), $T_{\rm bg}$=20~K model (middle
row) and $T_{\rm bg}$=30~K model (bottom panel). In particular, the black solid and 
red dashed line in the left column are 
the instantaneous mass accretion rate onto the star $\dot{M}_\ast$ and the time-averaged (over 15000~yr)
mass accretion rate onto the disk $\langle \dot{M}_{\rm d}\rangle$, respectively. 
The black solid and red dashed lines in the right column are the accretion 
and photospheric luminosities, respectively.

A comparison of the three models reveals that the $T_{\rm bg}$=20~K model exhibits 
a vigorous burst activity comparable in strength and frequency to that of the $T_{\rm bg}$=10~K model. However, the 
duration of the burst phase appears to be shorter in the higher-$T_{\rm bg}$ model.
As we further increase the background temperature to $T_{\rm bg}$=30~K, the burst activity decreases
notably,  yet there are two well-defined accretion and luminosity outbursts that reveal 
the disk is still prone to fragmentation. In fact, the magnitude of these bursts is much stronger
than in the lower-$T_{\rm bg}$ models, indicating that a higher background radiation favors
the formation of more massive fragments (though in a much smaller quantity).
As was expected from theoretical grounds, the photometric luminosity $L_{\rm\ast,ph}$ 
is greater in models with higher $T_{\rm bg}$, but so is the mass accretion rate onto 
the disk $\langle\dot{M}_{\rm d}\rangle$ (at least in the early phase). 
It appears that an elevated mass accretion 
rate onto the disk outweighs the stabilizing influence of the background and stellar irradiation.
We conclude that protostellar disks illuminated by the background irradiation with temperatures of the
order of 30~K (and probably higher) are still prone to fragmentation and development of the burst 
mode of accretion.

\begin{figure}
  \resizebox{\hsize}{!}{\includegraphics{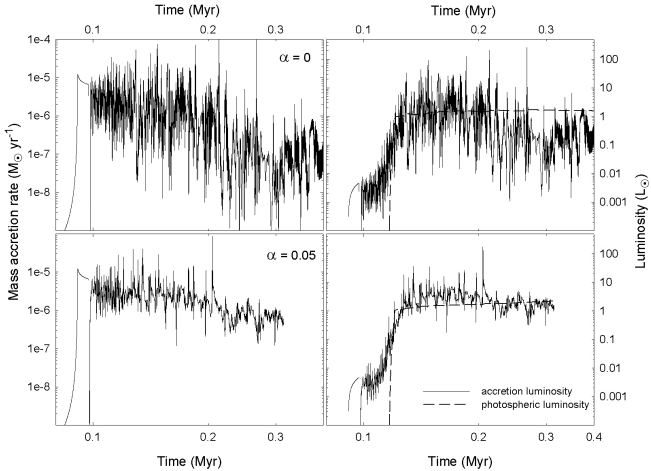}}
      \caption{Mass accretion rates onto the star (left column) and stellar luminosity (right column)
      as a function of time since the beginning of collapse
      in the $\alpha=0$ model (top) and $\alpha=0.05$ model (bottom). 
      In particular, the solid and dashed lines 
      in the right column present the accretion and photospheric luminosities, respectively. }
         \label{fig10}
\end{figure}

\subsection{The effect of viscosity}
\label{visceffect}
The effect of disk viscosity on the burst mode of accretion was studied by us in the context
of barotropic disks \citep{VB09a}. For the usual $\alpha$-parameterization of \citet{SS73} and
temporally and spatially constant $\alpha$, disks with lower values of $\alpha$ are disposed to
stronger fragmentation and demonstrate a stronger burst mode of accretion. In addition,
the accretion variability also increases along the line of decreasing $\alpha$ \citep{VB09a,Vor09}.

In all models considered so far, we have adopted $\alpha=0.005$. 
To see how different values of $\alpha$ could affect our conclusions, we run two models with 
$\alpha=0$ and $\alpha=0.05$ 
but other parameters identical to those of the reference model (see Table~\ref{table1}). 
Figure~\ref{fig10} presents the mass accretion rates onto the star (left column) and luminosities (right
column) in the $\alpha=0$ model (top row) and $\alpha=0.05$ model (bottom row).  
As expected, the $\alpha=0$ model demonstrates a vigorous burst activity, while the $\alpha=0.05$ model
shows only one strong luminosity outburst with $L_{\rm accr}$
in excess of 100~$L_\sun$, with other outbursts characterized by $L_{\rm accr}\la20~L_\sun$.
We confirm that a factor of 10 increase in $\alpha$ does not suppress 
disk fragmentation completely. However, an additional strong source of mass transport via 
viscous torques reduces the disk mass and this acts to moderate
the disk propensity to fragment. 

The problem with the $\alpha=0.05$ model is that it demonstrates
hardly any accretion episodes with $\dot{M}_\ast <10^{-6}~M_\sun$~yr$^{-1}$ in the early 0.2~Myr 
of evolution. The lack of low-rate accretion in disks with $\alpha>0.01$
was also found in the context of barotropic disks \citep{VB09a,Vor09} and this
confronts recent observations of \citet{Enoch09}, who find that a considerable 
fraction of Class~I sources in young star-forming regions have inferred accretion rates 
below $10^{-6}~M_\sun$~yr$^{-1}$. We note that both the $\alpha=0$ and $\alpha=0.005$
models show plenty of such low-accretion episodes. We therefore argue that disk viscosity in the 
embedded 
phase is unlikely to be characterized by $\alpha \gtrsim 0.01$.

In a broader context of viscous ($\alpha<0.01$) versus non-viscous ($\alpha=0$) models,
the former seem to yield accretion rates in the Class~II (or T Tauri) phase that
are a factor 2--3 greater than those of the non-viscous model \citep{VB07,VB08}. As a result, an addition
of $\alpha$-transport helps to bring Class~II disk masses in better agreement with observations \citep{Vor09a}.
On the other hand, the early disk evolution (Class~0 and I phases) is weakly affected by $\alpha$-viscosity
because mass and angular momentum transport in this stage is largely dominated by gravitational torques
\citep{VB09a}. This can also be seen from the comparison of Figure~\ref{fig5} with the top panels 
of Figure~\ref{fig10}---there is little {\it qualitative} difference in the mass accretion history 
between the $\alpha=0$ and $\alpha=0.005$ models in the EPSF.

%

\begin{figure}
  \resizebox{\hsize}{!}{\includegraphics{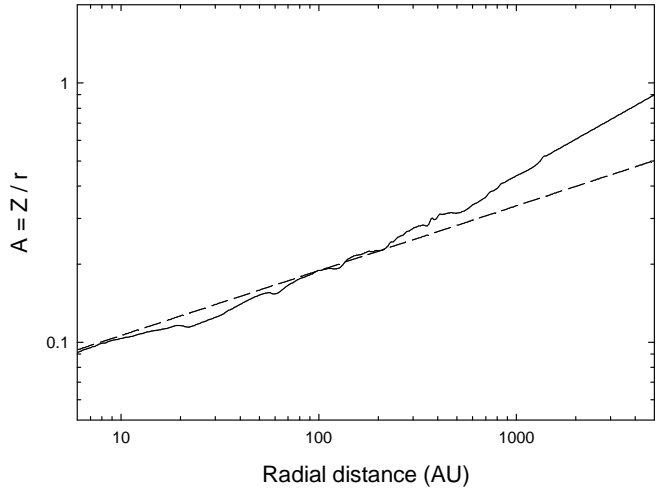}}
      \caption{Aspect ratio $A$ of the disk vertical scale height to radius ($Z/r$) as a
function of radius ($r$). The thick solid line shows $A$ in the reference model at $t=0.19$~Myr 
after the formation of the central star. The dashed line presents the aspect ratio 
derived from the following expression $Z/r= 0.1 (r/100~\mathrm{AU})^{0.25}$, 
as suggested by detailed disk vertical structure modeling by \citet{DAlessio99}.}
         \label{fig11}
\end{figure}

\section{Model limitations}
\label{discuss}
In this section, we discuss several assumptions in our model that can potentially influence 
our results. 
{\it 1) The onset of photospheric luminosity.}
As was discussed in Section~\ref{energy}, the stellar age in D'Antona \& Mazitelli's (1997) 
pre-main 
sequence evolution tracks is difficult to relate to the physical evolution 
time in numerical simulation of cloud core collapse.  We have equated their zero-point 
time to the time when the second atomic core
presumably starts to form in our numerical simulations. This may be a {\it conservative} assumption.
In D'Antona \& Mazitelli models, the evolution generally begins from a central temperature
of $\log(T_{\rm c})=5.7$, i.e., at a time instance just preceding deuterium burning, 
and it may take some time for the forming second core to ignite deuterium burning
in its interior. Hence, the photospheric luminosity may turn on somewhat later
than assumed in our numerical simulations  and this
could actually act to increase the disk susceptibility to fragmentation. \\
{\it 2) Accretion rate onto the star}. In our models, the size of the sink cell $r_{\rm sc}$=6~AU 
is larger than the stellar radius. The inner disk at $r < 6$~AU may add additional
variability to the accretion rates, in particular due to the thermal
instability \citep{Bell94} or magneto-rotational instability 
\citep{Armitage01,Zhu09}. These effects may somewhat 
alter the temporal behavior of the actual accretion rates onto the stellar surface 
and affect the accretion luminosity. However, our accretion rates are in good accord with those
inferred for nearby star-forming regions  and  we believe that the actual accretion
rates onto the stellar surface are not substantially different from those calculated in our
modeling. \\
{\it 3) Jet efficiency}. Protostellar jets may evacuate a substantial fraction of the accreting 
material, reducing the effective mass of the star as compared to the case without jets. 
In our modeling, we set the jet efficiency to 10\%, which means that the stellar mass
is systematically lower by 10\% than that of the non-jet case. This in fact promotes gravitational
instability in the disk as the disk-to-star mass ratio is increased accordingly. However,
the jet efficiency may be higher and amount to 30\% and possibly more 
\citep[e.g.,][]{Shu99}.
This would act to further destabilize the disk. \\
{\it 4) Stellar wobbling}. The position of the central star in our models is fixed in the coordinate
center. However, the star may move in response to the non-axisymmetric gravitational field of the disk.
Semi-analytic considerations suggest that this stellar wobbling may amplify  
gravitational instability in the disk \citep{Adams89}, though the recent numerical hydrodynamics 
simulations find this effect insignificant \citep{Durisen10}. 
 To implement such a mechanism in our models
is however not easy due to the presence of singularity in the coordinate center on the polar grid. 
We plan to explore the effect of stellar wobbling in a future study. \\
{\it 5) Binary or multiple system formation.}
In addition to the clump formation that we see in the present models, 
we have also seen the formation of a binary companion (or multiple companions)
in the outer disk in models with $M_{\rm cl}\ga 1.7~M_\sun$ 
and $\beta\ga 2.0\times 10^{-2}$. These models will be presented in a future paper. 
We note that to fully capture binary formation in the outer regions with our logarithmic grid, 
we need even higher numerical resolution than in the present study.  \\
{\it 5) Magnetic fields.}
Frozen-in magnetic fields moderate the burst activity due to an effective increase 
in the $Q$-parameter \citep{VB06}. For a spatially and temporally uniform 
mass-to-flux ratio, the magnetic tension acts as a simple dilution of gravity, thus 
effectively lowering the disk surface density, and
the magnetic pressure is a multiple of the gas pressure, thus providing an effective increase to
the disk sound speed. A more comprehensive study of the effect of magnetic fields,
including ambipolar diffusion and magnetic braking, is planned for a future paper.

\section{Conclusions}
We have revisited our original results on the burst mode of accretion \citep{VB05,VB06}, 
paying special attention to the thermal processes in protostellar disks around low-mass protostars.
Our new model takes into account radiative cooling from
the disk surface, viscous and shock heating, and also stellar and background irradiation.
Thanks to the use of the thin-disk approximation, we can run uninterrupted numerical hydrodynamics 
simulations from the prestellar phase to the early T Tauri phase, fully capturing the embedded phase
of star formation (EPSF). We find the following.


\begin{itemize}
\item The EPSF is likely the only episode of disk evolution when disk 
fragmentation can take place. However, disk susceptibility to fragmentation in this phase depends 
crucially on the {\it initial conditions} in a natal cloud core. 

\item Higher initial core angular momentum and mass lead to the formation of more massive 
and extended disks and, therefore, favor disk fragmentation. On the other hand, 
a higher temperature of the background irradiation $T_{\rm bg}$ may moderate 
the disk propensity to fragment. In particular, higher $T_{\rm bg}$ appears to favor 
the formation of more massive fragments though in much fewer numbers.

\item A higher rate of mass infall onto the disk than that onto the star in the EPSF
does not guarantee disk fragmentation if the disk is not sufficiently large and massive. 

\item For disk fragmentation to occur, {\it both} the Toomre $Q$-parameter and 
${\cal G}$-parameter (ratio of the local cooling time to the dynamical time) must be below some 
critical value, taken to be unity in this paper, confirming many previous studies 
on disk instability and fragmentation. 


\item Most (but possibly not all) fragments that from in the EPSF are driven into the inner disk 
regions and probably onto the star, triggering 
mass accretion and luminosity bursts similar in magnitude to those of the FU-Orionis-type
and EX-Lupi-like stars. This burst mode of accretion is a robust phenomenon that is expected
to exist in a variety of environments and for a variety of systems with different physical
properties. The intensity of the burst mode correlates with the disk propensity to fragment. 

\item Fragmenting disks drive highly variable accretion rates onto the star ranging from 
$10^{-8}~M_\sun$~yr$^{-1}$ to $10^{-4}~M_\sun$~yr$^{-1}$.
Protostellar disks that are gravitationally unstable but stable to fragmentation
are characterized by a considerably weaker accretion variability with only an 
order of magnitude flickering.

\item The intensity of the burst mode of accretion is sensitive to the amount 
of $\alpha$-viscosity present 
in protostellar disks and appears to subside with increasing $\alpha$. The lack of 
strong variability in disks with a spatially and temporally uniform $\alpha\ga 0.05$
contradicts observations \citep[e.g.][]{Enoch09} and renders such disk not viable.


\end{itemize}

\acknowledgments    
   E.I.V. gratefully acknowledges present support 
   from an ACEnet Fellowship, RFBR grant 10-02-00278, and Ministry of Education grant 
   RNP 2.1.1/1937. Numerical simulations were done 
   on the Atlantic Computational Excellence Network (ACEnet),
   on the Shared Hierarchical  Academic Research Computing Network (SHARCNET),
   and at the Center of Collective Supercomputer
   Resources, Taganrog Technological Institute at Southern Federal University.
   S. B. was supported by a Discovery Grant from the Natural Sciences and 
   Engineering Research Council of Canada. 

\appendix
\section{The thin-disk approximation}
\label{thindisk}
The thin-disk approximation is well justified as long as the aspect ratio $A=Z/r$ of the
disk vertical scale height $Z$ to radius $r$ does not considerably exceed
0.1. In a Keplerian disk, $Z=c_{\rm s}/\Omega$ and noticing that the angular velocity
is $\Omega=(G M_\ast/r^3)^{1/2}$ and the sound speed is $c_{\rm s}\le Q_{cr} \pi G \Sigma/\Omega$,
the aspect ratio can be expressed as
\begin{equation}
A \le { Q_{\rm cr} \, M_{\rm d}(r) \over C M_\ast},
\end{equation}
where $M_{\rm d}(r)=\int\Sigma(r,\phi)\, r \,dr \,d\phi$ is the disk mass contained within radius 
$r$, $M_\ast$ is the mass of the central star,
$Q_{\rm cr}$ is the critical Toomre parameter, and $C$ is a constant, the actual value of which 
depends on the gas surface density distribution $\Sigma$ in the disk. 
For a disk of constant surface density,
$C$ is equal unity. However, circumstellar disks are characterized by
surface density profiles declining with radius. For the scaling
$\Sigma \propto r^{-1.5}$ typical for our disks, $C=4$. 
Adopting further $Q_{\rm cr}=2$ and $M_{\rm d}(r)/M_\ast=0.5$, 
which are typical {\it upper} limits in our numerical simulations, we obtain 
$A\la0.25$. This analysis demonstrates that the thin-disk approximation 
is certainly valid in the inner regions where $M_{\rm d}(r)/M_\ast$ is small, but
may become only marginally valid at large $r$ where $M_{\rm d}(r)/M_\ast$ would approach
its maximum value.  

The azimuthally-averaged radial distribution of the aspect ratio $A=Z/r$ in the reference model at $t=0.19$~Myr
after the formation of the central star in shown by the solid line Figure~\ref{fig11}. The vertical
scale height $Z$ is calculated assuming a local vertical hydrostatic equilibrium in the disk using 
the method described in \citet{VB09a}. Figure~\ref{fig11} reinforces our analytical estimates and demonstrates
the thin-disk approximation is certainly obeyed in the disk. Our disks rarely exceed 1000~AU in
radius and the corresponding aspect ratio is kept in the 0.1--0.4 limits. Only at radial distances well
in excess of 1000~AU may the thin-disk approximation be violated.

\section{Radiation flux from the central star}
\label{irradiation}
In order to calculate the radiation flux from the central
star $F_{\rm irr}$ at a given radial distance $r$ using Eq.~(\ref{fluxCS}),
one needs to know the incidence angle of radiation arriving at the
disk surface $\gamma_{\rm irr}$ (i.e., the angle between the light rays and the perpendicular to
the disk surface).
For a flaring disk, the cosine of $\gamma_{\rm irr}$ can be expressed as 
\begin{equation}
\cos\gamma_{\rm irr}=\cos\alpha_{\rm irr} \cos\beta_{\rm irr} \left( \tan\alpha_{\rm irr}
- \tan\beta_{\rm irr} \right),
\end{equation}
where $\cos\alpha_{\rm irr}=dr/(dr^2+dZ^2)^{1/2}$, $\cos\beta_{\rm irr}=r/(r^2+Z^2)^{1/2}$,
$\tan\alpha_{\rm irr}=dZ/dr$, and $\tan\beta_{\rm irr}=Z/r$. In most cases, $\cos\alpha_{\rm irr}\approx
1$ and $\cos\beta_{\rm irr}\approx 1$, since $Z/r \ll 1$ (thin disk) and $dZ/dr \ll 1$ (weak flaring).
Nevertheless, we use the complete expression for $\cos\gamma_{\rm irr}$.

In the reality, the disk surface may not always be of the concave shape, so that $\tan\alpha_{\rm irr}-
\tan\beta_{\rm irr}>0$ and $F_{\rm irr}>0$. If the shape of the disk surface becomes convex, i.e., $\tan\alpha_{\rm irr}-
\tan\beta_{\rm irr}<0$, the irradiation flux $F_{\rm irr}$ becomes {\rm negative}.
Physically, this corresponds to the situation when part of the disk surface is shielded from the 
incoming radiation from the central star, for instance, by a local puffing of the disk. 
This may be potentially an important phenomenon. However,
taking this effect into account self-consistently may require the use of full radiation transfer using
ray tracing and is out of scope of the present paper. Therefore, to avoid this complication, we 
make use of the detailed vertical structure models of irradiated accretion disks around T Tauri 
stars by 
\citet{DAlessio99}. From their figure 1(b) (dashed curve) we have derived the following expression
$Z/r= 0.1 (r/100~\mathrm{AU})^{0.25}$ for the aspect ratio $Z/r$ as a function of radial distance $r$,
where $0.1$ is the ratio $Z/r$ at $r=100$~AU and exponent $0.25$ determines the degree of disk flaring
(for positive/negative exponents, the disk surface is concave/convex, respectively).
We adopt this relation with a modification according to our model, i.e., we actually calculate 
the aspect ratio $Z/r$ at $100$~AU using the azimuthally averaged value of the vertical scale height
$Z$. This would allow us to dynamically adjust the aspect ratio $Z/r$ according to the actual 
disk thickness but keep the disk shape concave throughout the simulation.

Another effect that has to be taken into account is the attenuation of stellar irradiation
by the infalling envelope in the embedded phase of star formation (EPSF). This is done by 
introducing 
a factor $A_{\rm irr}$ in Eq.~(\ref{fluxCS}) calculated as $A_{\rm irr}=M_{\rm cl}/(M_{\rm env}+M_{\rm
cl})$, where $M_{\rm cl}$ is the cloud core mass (stays fixed) 
and $M_{\rm env}$ is the gradually decreasing envelope mass. In the early EPSF, $M_{\rm cl}\approx M_{\rm env}$ and $A_{\rm
irr}\approx 0.5$, while in the late EPSF, $M_{\rm env} \rightarrow 0$ and $A_{\rm irr}\rightarrow 1$.

\section{Supplementary mathematical formula}
For the convenience of the reader and for completeness, we provide the actual 
expressions for $(\nabla \cdot {\bl \Pi})_p$, $(\nabla{\bl v})_{pp}$, and 
$\left[ \nabla \cdot \left( \Sigma {\bl v}_p \otimes {\bl v}_p \right)\right]_p$ used in our paper.
The components of $\nabla \cdot {\bl \Pi}$ in polar coordinates ($r,\phi$) are
\begin{eqnarray}
\label{divergence1}
\left( \nabla \cdot {\bl \Pi} \right)_r &=& {1\over r} {\partial \over \partial r} r \Pi_{rr} +
{1 \over r}  {\partial \over \partial \phi} \Pi_{r\phi} - {\Pi_{\phi\phi} \over r}, \\
\label{divergence2}
\left( \nabla \cdot {\bl \Pi} \right)_\phi &=& {\partial \over \partial r} \Pi_{\phi r}
+ {1\over r} {\partial \over \partial \phi} \Pi_{\phi\phi} + 2 {\Pi_{r\phi} \over r},
\end{eqnarray}
where we have neglected the contribution from off-diagonal components $\Pi_{rz}$ and $\Pi_{\phi z}$.
The components of the viscous stress tensor $\bl \Pi$ in polar coordinates ($r,\phi$) can be found
from Eq.~(\ref{stressT}) according to the usual rules.

When calculating the symmetrized velocity gradient tensor $\nabla {\bl v}$, only the
following planar components are assumed to be non-zero:
\begin{eqnarray}
\left( \nabla {\bl v} \right)_{rr} & = & {\partial v_r \over \partial r}, \\
\left( \nabla {\bl v}  \right)_{r\phi} & = & {1\over r} {\partial v_r \over \partial \phi} +
{\partial v_\phi \over \partial r} - {u_\phi \over r},  \\ 
\left( \nabla {\bl v} \right)_{\phi\phi} & = & {1\over r} {\partial v_\phi \over \partial \phi} + {v_r
\over r}. 
\end{eqnarray}
The symmetric dyadic $\Sigma {\bl v}_p \otimes {\bl v}_p$ is a rank-two tensor expressed in polar coordinates
($r,\phi$) as
\begin{equation}
\Sigma {\bl v}_p \otimes {\bl v}_p = 
\left| 
\begin{array}{ll}
\Sigma v_r v_r & \Sigma v_r v_\phi \\
\Sigma v_\phi v_r & \Sigma v_\phi v_\phi
\end{array}
\right|.
\end{equation}
The planar components of $\left[ \nabla \cdot \left( \Sigma {\bl v}_p \otimes {\bl v}_p \right)\right]_p$
can then be found using Eqs.~(\ref{divergence1}) and (\ref{divergence2}) 
with ${\bl \Pi}$ substituted by $\Sigma {\bl v}_p \otimes {\bl v}_p$.

Finally, the viscous heating term $\left(\nabla \bl{v}\right)_{pp^\prime}:\Pi_{pp^\prime}$ in the energy
balance equation is the convolution of two rank-two tensors and its expression in the thin-disk approximation
(neglecting the off-diagonal components) is as follows
\begin{equation}
\left(\nabla \bl{v}\right)_{pp^\prime}:\Pi_{pp^\prime}={2 \mu \over 3} \left\{\left(\nabla \bl{v}\right)_{rr}^2 + 
\left(\nabla \bl{v}\right)_{\phi\phi}^2 +\left[ \left(\nabla \bl{v} \right)_{rr} -  
\left(\nabla \bl{v} \right)_{\phi\phi} \right]^2 \right\} + 2\left(\nabla \bl{v} \right)_{r\phi} \Pi_{r\phi}.
\end{equation}

\end{document}